\documentclass[twocolumn]{aastex631}
\shorttitle{Non-Thermal Velocity and Magnetic Energy in Flares}
\shortauthors{McKevitt et al.}

\usepackage[acronym]{glossaries} 
\usepackage{savesym}
\savesymbol{tablenum}
\usepackage{siunitx} 
\restoresymbol{SIX}{tablenum}
\DeclareSIUnit\erg{erg} 
\DeclareSIUnit\angstrom{\text{Å}}
\DeclareSIUnit\arcsec{\text{arcsec}} 
\DeclareSIUnit\Gauss{\text{G}}
\DeclareSIUnit\pixel{\text{pix}}
\glsdisablehyper 
\newacronym{euv}{EUV}{extreme ultraviolet}
\newacronym{cme}{CME}{coronal mass ejection}
\newacronym{nlfff}{NLFFF}{non-linear force-free field}
\newacronym[first={Helioseismic and Magnetic Imager \protect\citep[HMI;][]{Scherrer2012TheSDO}}]{hmi}{HMI}{Helioseismic and Magnetic Imager}
\newacronym[first={\gls{euv} Imaging Spectrometer \protect\citep[EIS;][]{Culhane2007TheHinode}}]{eis}{EIS}{\gls{euv} Imaging Spectrometer}
\newacronym{sep}{SEP}{solar energetic particle}
\newacronym[first={Atmospheric Imaging Assembly \protect\citep[AIA;][]{Lemen2012TheSDO}}]{aia}{AIA}{Atmospheric Imaging Assembly}
\newacronym[first={Solar Dynamics Observatory \protect\citep[SDO;][]{Pesnell2012TheSDO}}]{sdo}{SDO}{Solar Dynamics Observatory}
\newacronym[first={Geostationary Operational Environmental Satellite \protect\citep[GOES;][]{Landecker2009GOESC}}]{goes}{GOES}{Geostationary Operational Environmental Satellite}
\newacronym{xrs}{XRS}{X-ray Sensor}
\newacronym[longplural=Space-weather HMI Active Region Patches]{sharp}{SHARP}{Space-weather HMI Active Region Patch}
\newacronym{harp}{HARP}{HMI Active Region Patch}
\newacronym[first={cylindrical equal area \protect\citep[CEA;][]{Bobra2014ThePatches}}]{cea}{CEA}{cylindrical equal area}
\newacronym[first={physics-informed neural network \protect\citep[PINN;][]{Raissi2019Physics-informedEquations}}]{pinn}{PINN}{physics-informed neural network}
\newacronym{gpu}{GPU}{graphics processing unit}
\newacronym[first={the EIS Python Analysis Code \protect\citep[EISPAC;][]{Weberg2023EISPACCode}}]{eispac}{EISPAC}{the EIS Python Analysis Code}

\begin{document}

\title{The Link Between Non-Thermal Velocity and Free Magnetic Energy in Solar Flares}

\correspondingauthor{James E. McKevitt}
\email{james.mckevitt.21@ucl.ac.uk}

\author[0000-0002-4071-5727]{James McKevitt}
\affiliation{University College London, Mullard Space Science Laboratory \\
Holmbury St Mary, Dorking \\
Surrey, RH5 6NT, UK}
\affiliation{University of Vienna, Institute of Astrophysics \\
Türkenschanzstrasse 17 \\
Vienna 1180, Austria}

\author[0000-0002-9309-2981]{Robert Jarolim}
\affiliation{University of Graz, Institute of Physics \\ Universitätsplatz 5 \\
Graz 8010, Austria}

\author[0000-0001-9346-8179]{Sarah Matthews}
\affiliation{University College London, Mullard Space Science Laboratory \\
Holmbury St Mary, Dorking \\
Surrey, RH5 6NT, UK}

\author[0000-0002-0665-2355]{Deborah Baker}
\affiliation{University College London, Mullard Space Science Laboratory \\
Holmbury St Mary, Dorking \\
Surrey, RH5 6NT, UK}

\author[0000-0003-4867-7558]{Manuela Temmer}
\affiliation{University of Graz, Institute of Physics \\ Universitätsplatz 5 \\
Graz 8010, Austria}

\author[0000-0003-2073-002X]{Astrid Veronig}
\affiliation{University of Graz, Institute of Physics \\ Universitätsplatz 5 \\
Graz 8010, Austria}

\author[0000-0002-6287-3494]{Hamish Reid}
\affiliation{University College London, Mullard Space Science Laboratory \\
Holmbury St Mary, Dorking \\
Surrey, RH5 6NT, UK}

\author[0000-0002-0053-4876]{Lucie Green}
\affiliation{University College London, Mullard Space Science Laboratory \\
Holmbury St Mary, Dorking \\
Surrey, RH5 6NT, UK}

\begin{abstract}

The cause of excess spectral line broadening (non-thermal velocity) is not definitively known, but given its rise before and during flaring, the causal processes hold clues to understanding the triggers for the onset of reconnection and the release of free magnetic energy from the coronal magnetic field. A comparison of data during a 9-hour period from the extreme ultraviolet (EUV) Imaging Spectrometer (EIS) on the Hinode spacecraft~-~at a 3-minute cadence~-~and non-linear force-free field (NLFFF) extrapolations performed on Helioseismic and Magnetic Imager (HMI) magnetograms~-~at a 12-minute cadence~-~shows an inverse relationship between non-thermal velocity and free magnetic energy on short timescales during two X-class solar flares on 6 September 2017. Analysis of these results supports suggestions that unresolved Doppler flows do not solely cause non-thermal broadening and instead other mechanisms like Alfvén wave propagation and isotropic turbulence have a greater influence.

\end{abstract}
\glsresetall 


\section{Introduction}

Solar flares are widely believed to occur as the result of the sudden and impulsive release of energy stored in non-potential magnetic fields \citep{Toriumi2019Flare-productiveRegions,Priest2002TheFlares}. These fields, rather than following the lowest energy configuration, exhibit a degree of twist or shear. The energy difference between these non-potential, i.e. current-carrying fields and their lowest energy state represents the energy stored in the magnetic field. It is known as free magnetic energy and is available to produce flares and \glspl{cme} \citep{Wiegelmann2021SolarFields}. When a non-potential field transitions to a lower energy state through magnetic reconnection, the stored energy is released into the solar atmosphere. This process, in line with the standard flare model \citep{Shibata2011SolarProcesses}, leads to plasma heating and particle acceleration \citep{Pontin2022MagneticModelling,Fletcher2011AnFlares,Benz2017FlareObservations}, serving as the fundamental mechanism for energy release in solar flares and the ejection of material in \glspl{cme}.

The creation of non-potential fields is primarily attributed to the emergence of magnetic flux~-~the ascent of twisted and distorted bundles of magnetic field lines through the solar convective zone, culminating in their emergence through the photosphere as twisted flux tubes~-~and their interaction with pre-existing fields \citep{Leka1996EvidenceFlux}. Other contributing factors include the shearing and twisting of the magnetic structures at the photosphere, for instance, the movement of footpoints \citep{Park2018PhotosphericOccurrence}.

The non-thermal broadening of \gls{euv} and soft X-ray lines has been observed to increase substantially during flaring~-~reaching velocities as high as \SI{200}{\kilo\meter\per\second} \citep[e.g.,][]{Doschek1980High-resolutionFlares}~-~and before flare onset, often showing enhancements tens of minutes prior to the start of the flare impulsive phase, \citep[e.g.,][]{Harra2001NonthermalFlare}. However, the precise relationship between the flare energy release processes and the origin of the excess line broadening remains unclear.

Active region NOAA~12673 has been the focus of significant attention in the solar community due to the notably energetic solar events it generated. First observed on the eastern solar limb on 31 August 2017, this region underwent substantial flux emergence starting on 3 September 2017. This rapid evolution precipitated the production of several M- and X-class flares, \glspl{cme}, and \gls{sep} events \citep[e.g.,][]{Yan2018Successive12673,Verma2018TheFlux}. X-class flares, being the most energetic, are the class of flare most clearly reflected in both the free magnetic energy and non-thermal velocity, and are therefore ideal to use to study the relationship between these parameters.

Our study focuses on active region NOAA~12673 during the time period from 06:00 to 14:48~UTC on 6 September 2017, an approximately 9-hour window that saw a confined X2.2 flare and an eruptive X9.3 flare \citep[e.g.,][]{Gupta2021MagneticFlares,Hou2018Eruption24,Mitra2018Successive6}. We undertook a comprehensive analysis of the free magnetic energy release in the active region during this interval, exploring its correlation with non-thermal velocity for multiple \gls{euv} emission lines. In this paper, we study the corona's response to the non-potential field configuration, particularly with respect to the coronal emission line widths, and quantify this relationship.

\section{Observations and extrapolations}

This study combines data taken by the \gls{eis} onboard the Hinode spacecraft \citep{Kosugi2007TheOverview} with photospheric magnetograms from the \gls{hmi} on board the \gls{sdo}. We additionally use the \gls{goes} system, specifically data from the X-ray Sensor (XRS) of the GOES-13 spacecraft.

\subsection{EUV Observations}

The \gls{euv} data used in this study were gathered between 06:20 and 14:50~UTC on 6 September 2017 by \gls{eis}. The \gls{eis} instrument is a scanning slit spectrometer that observes the solar corona and upper transition region in two \gls{euv} wavebands: \SIrange{170}{211}{\angstrom} and \SIrange{246}{292}{\angstrom} \citep{Culhane2007TheHinode,Young2007EUVHinode/EIS}.

We analysed a series of raster scans taken during the observation period while \gls{eis} was operating a high-cadence, reduced field-of-view flare study (study 473\footnote{\href{https://solarb.mssl.ucl.ac.uk/SolarB/}{https://solarb.mssl.ucl.ac.uk/SolarB/}}) to capture the response of the coronal plasma to a flare. This study completes repeated scanning rasters, each with 30 pointing positions taken sequentially from west to east and with a scan step size of \SI{4}{\arcsec}. The exposure time for each pointing position was approximately 4~seconds which, when considering other instrumental operations, results in a raster cadence of approximately 3~minutes. The time given for each observation henceforth refers to the midpoint of the observation. There are a number of emission lines observed in this configuration ranging from cool to flaring lines. We focused on the strong \ion{Fe}{14}~\SI{264.79}{\angstrom} and \ion{Fe}{14}~\SI{274.20}{\angstrom}  ($\log~T_{\text{max}}=6.3$) coronal lines given they demonstrated a high signal/noise during both X-class flares, did not saturate during the peak intensities during flaring, and allowed consistently good fits throughout the detector and throughout the time series. The former is recommended for probing hotter parts of active regions, and so can be expected to react quickly to flaring activity. It is also not overlapped by any other known emission lines. The latter is blended with a small contribution from \ion{Si}{7}~\SI{275.35}{\angstrom} which can be safely neglected \citep{Young2007EUVHinode/EIS}. These two lines were used in tandem, with the latter used to verify the response of the former, and indeed in our study, they both demonstrate very similar results. The strong \ion{Ca}{17}~\SI{192.82}{\angstrom} ($\log~T_{\text{max}}=6.7$) flaring line was also used given its more specific response to flares, but required special additional consideration as is discussed later.

\begin{figure*}
  \gridline{
    \fig{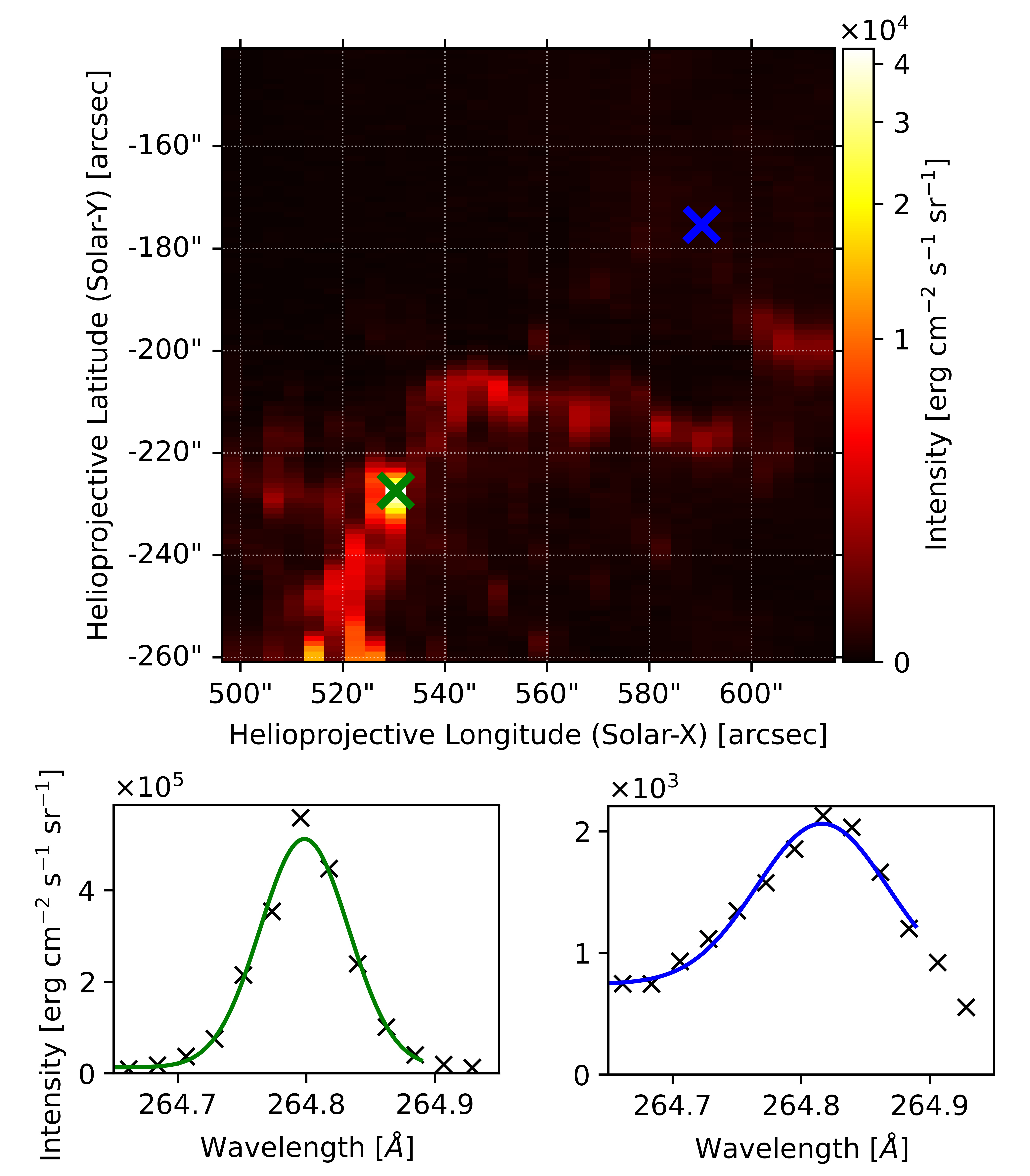}{0.45\textwidth}{(a) \ion{Fe}{14}~\SI{264.79}{\angstrom} intensity map and sample fitted spectra for the highest intensity (green) and median intensity (blue) pixels.}
    \fig{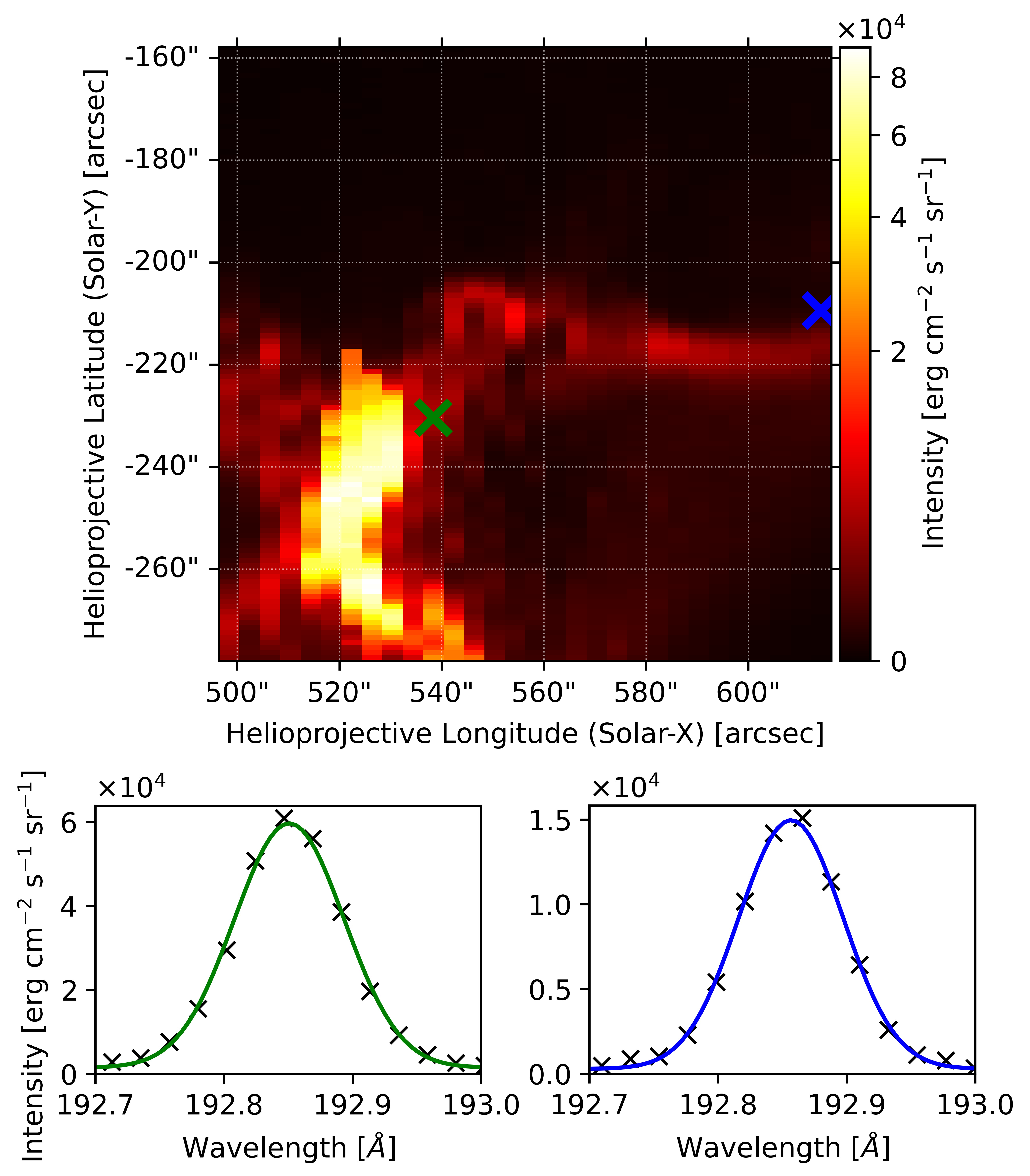}{0.45\textwidth}{(b) \ion{Ca}{17}~\SI{192.82}{\angstrom} intensity map and sample fitted spectra for the 85\% intensity (green) and median intensity (blue) pixels. Details on the saturation of some pixels are in the text.}
  }
  \caption{Intensity map and sample fitted spectra for active region NOAA~12673, using data from Hinode/EIS at 11:58~UTC on 6 September 2017 during the onset of the X9.3 flare.}
  \label{fig:fits}
\end{figure*}

Gaussian fits were carried out for these lines, for each pixel in each raster scan using \gls{eispac}. An example of this fitting can be seen in Figure~\ref{fig:fits}. This fitting procedure results in maps of the line width which were converted into non-thermal velocity ($v_{nt}$) using

\begin{equation}
    {\text{FWHM}_{o}}^2={\text{FWHM}_{i}}^2+4\ln{2}\left(\frac{\lambda}{c}\right)^2\left({v_t}^2+{v_{nt}}^2\right),
\end{equation}

where $\text{FWHM}_{o}$ and $\text{FWHM}_{i}$ refer to the observed and instrumental full width at half maximum values respectively, and where $\lambda$, $v_t$ and $c$ refer to the central wavelength of the fitted Gaussian, the associated thermal velocity, and the speed of light.

The uncertainty in the instrumental width can be combined with the statistical error in the fitted Gaussian~-~partly caused by uncertainty in the measurement of each point in the emission line~-~using standard error propagation \citep[e.g.,][]{Bevington2003DataSciences} to estimate the error in non-thermal velocity measurements to be approximately 25\%.

The values for non-thermal velocity as observed by each pixel in the image, such as shown in Figure~\ref{fig:fits}, were then averaged for each observation to generate one non-thermal velocity value for each observation time, resulting in a time series. This was found to be the most effective way to include a sufficient sample size in each observation to reduce noise, while still clearly capturing increases in non-thermal velocity.

As aforementioned, \ion{Ca}{17}~\SI{192.82}{\angstrom} requires careful consideration given it is part of a complex blend comprising seven other known lines. During low periods of activity, various methods can be used to estimate their respective contributions and isolate the \ion{Ca}{17} emission. However, during flaring \ion{Ca}{17} completely dominates and so any other contributions can be considered negligible \citep{Young2007EUVHinode/EIS,Ko2009HotHinode}. Additionally, several wavelength bins within some pixels became saturated during both X-class flares in this study (at \SI{5e5}{\erg\per\cm\squared\per\s\per\steradian}). It was determined that the best approach to address this was to exclude any pixels within which any of the wavelength bins had become saturated, these accounting for only 1\% of \ion{Ca}{17}~\SI{192.82}{\angstrom} pixels in the complete dataset and peaking at approximately 10\% of the \ion{Ca}{17}~\SI{192.82}{\angstrom} pixels in one observation during the X9.3 flare. This processing means that the \ion{Ca}{17}~\SI{192.82}{\angstrom} non-thermal velocity values should be considered a lower estimate.

The pointing information of the \gls{eis} data was corrected by co-aligning the \ion{Fe}{14}~\SI{264.79}{\angstrom} intensity maps with imaging performed at \SI{171}{\angstrom} by the \gls{aia} onboard \gls{sdo}. This required only small longitudinal corrections, increasing due to drift during the 9-hour period to just less than \SI{10}{\arcsec} by the end of the observation.

\subsection{Magnetic field observations}


In addition to the \gls{eis} data we also used data gathered by \gls{hmi} between 06:00 and 14:48~UTC on 6 September 2017, excluding a data gap between 06:12 and 08:24~UTC inclusive due to instrument downtime. The \gls{hmi} instrument generates full-disk vector photospheric magnetograms with a cadence of 12~minutes and at a resolution of about \SI{1}{\arcsec}, with a noise level of about \SI{100}{\Gauss} \citep{Hoeksema2014ThePerformance}.

For this study, we used the \gls{hmi} vector magnetograms to perform \gls{nlfff} extrapolations of the photospheric magnetic field to then subsequently calculate the free magnetic energy of the magnetic field. The magnetograms used were those provided by \glspl{sharp} in \gls{cea} projection, using \gls{sharp} 7115.

Based on the photospheric magnetic field, we performed a magnetic field extrapolation using the method developed by \cite{Jarolim2023ProbingNetworks}, allowing for the production of a maximum cadence time series based on magnetograms with high spatial resolution. Here, a \gls{pinn} is used to solve the force-free equation

\begin{equation}
    J\times{}B=0
\end{equation}

\noindent{}and divergence-free equation

\begin{equation}
    \nabla\cdot{}B=0,
\end{equation}

\noindent{}where $J$ is the electric current density and $B$ is the magnetic field, for the given boundary condition. We compute the magnetic field $B$ up to a height of approximately \SI{115}{\mega\meter} with re-binned magnetograms to 1/2 resolution, resulting in \SI{0.72}{\mega\meter\per\pixel}.

\begin{figure}
    \gridline{\fig{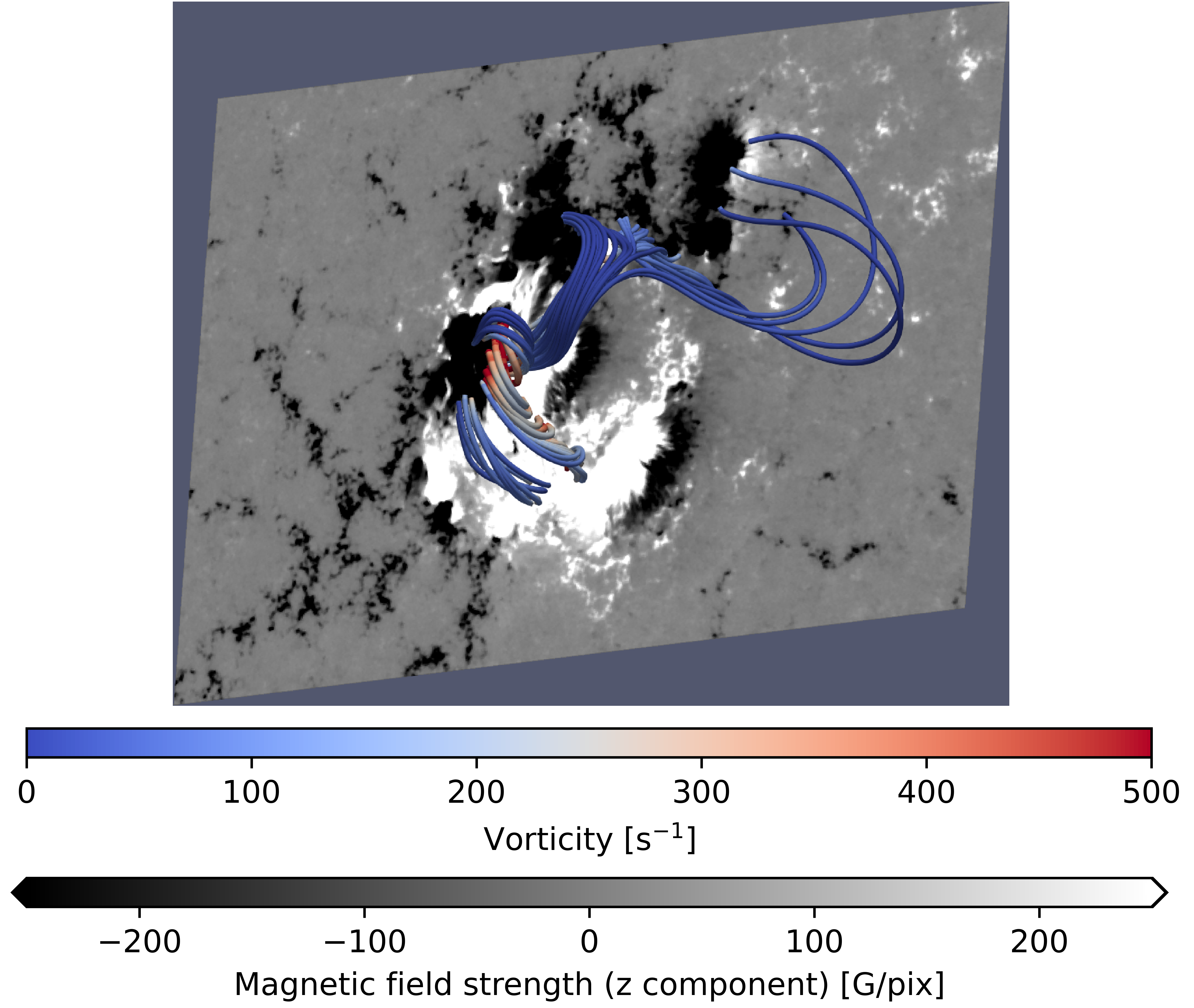}{.5\textwidth}{(a) Selected magnetic field lines as computed by the \acrshort{nlfff} modelling. The z~component of the magnetic field is shown at the bottom of the extrapolation.}}
    \gridline{\fig{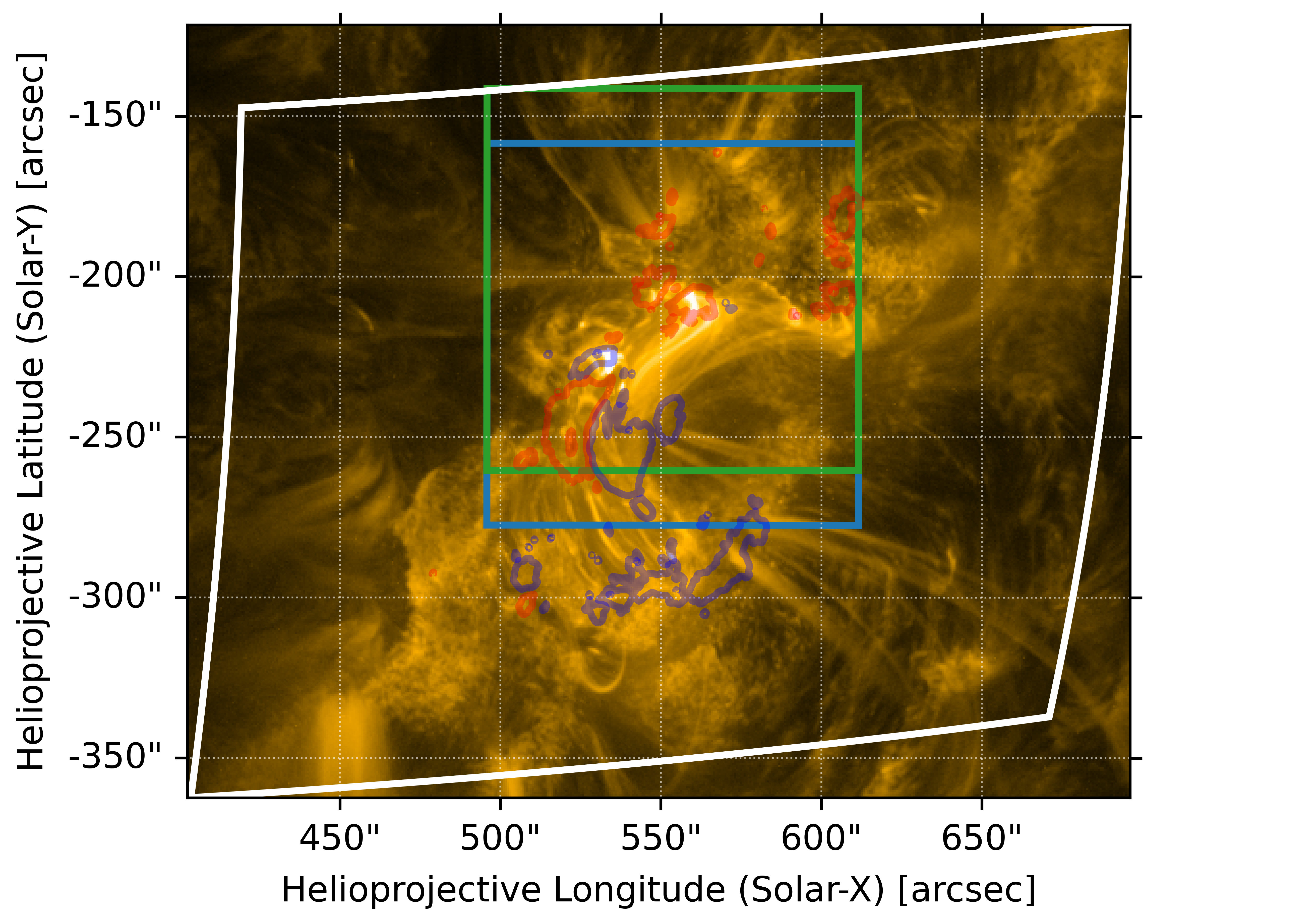}{.5\textwidth}{(b) AIA~\SI{171}{\angstrom} with overplotted contours of the photospheric negative (red) and positive (dark blue) magnetic field at -750~G and 750~G respectively. Also shown is the region of SHARP 7115 used for \acrshort{nlfff} extrapolations (white) and the \acrshort{eis} field of view for the \ion{Fe}{14}~\SI{264.79}{\angstrom} and \ion{Fe}{14}~\SI{274.20}{\angstrom} emission lines (green) and the \ion{Ca}{17}~\SI{192.82}{\angstrom} emission line (light blue).}}
    \caption{Cospatial maps of the \acrshort{nlfff} modelling and the coronal EUV emission showing the agreement of the magnetic structures resolved by the \acrshort{nlfff} extrapolation and the structures visible in the EUV. These plots are of the active region as observed at 11:36~UTC, just before the onset of the X9.3 flare.}
    \label{fig:mag_configs}
\end{figure}

Modelled field lines were validated against \gls{euv} emission structures imaged using AIA. Figure~\ref{fig:mag_configs} shows low-lying loops in the sheared core of the active region are well matched to structures seen in AIA~\SI{171}{\angstrom}.

The total magnetic energy within the simulation volume $V$ can then be calculated using

\begin{equation}
    E=\int_V{}\frac{B^2}{8\pi}dV.
\end{equation}

\noindent{}When the total magnetic energy is calculated for both the force-free and potential field, $E_{FF}$ and $E_{PF}$ respectively, the free magnetic energy can be estimated using

\begin{equation}
    E_{free}=E_{FF}-E_{PF}.
\end{equation}

\noindent{}To compute the potential field we use the approach by \cite{Sakurai1982GreensFields}. 

Furthermore, we calculate a column-integrated free magnetic energy, summed along the \gls{eis} line of sight, which allows the free magnetic energy to be spatially resolved for the active region in the same projection as \gls{eis}. This follows a similar procedure but integrates the total energies at an angle to the vertical to generate a two-dimensional map of the free magnetic energy.

\section{Results}

\begin{figure*}[ht!]
    \centering
    \includegraphics[width=\textwidth]{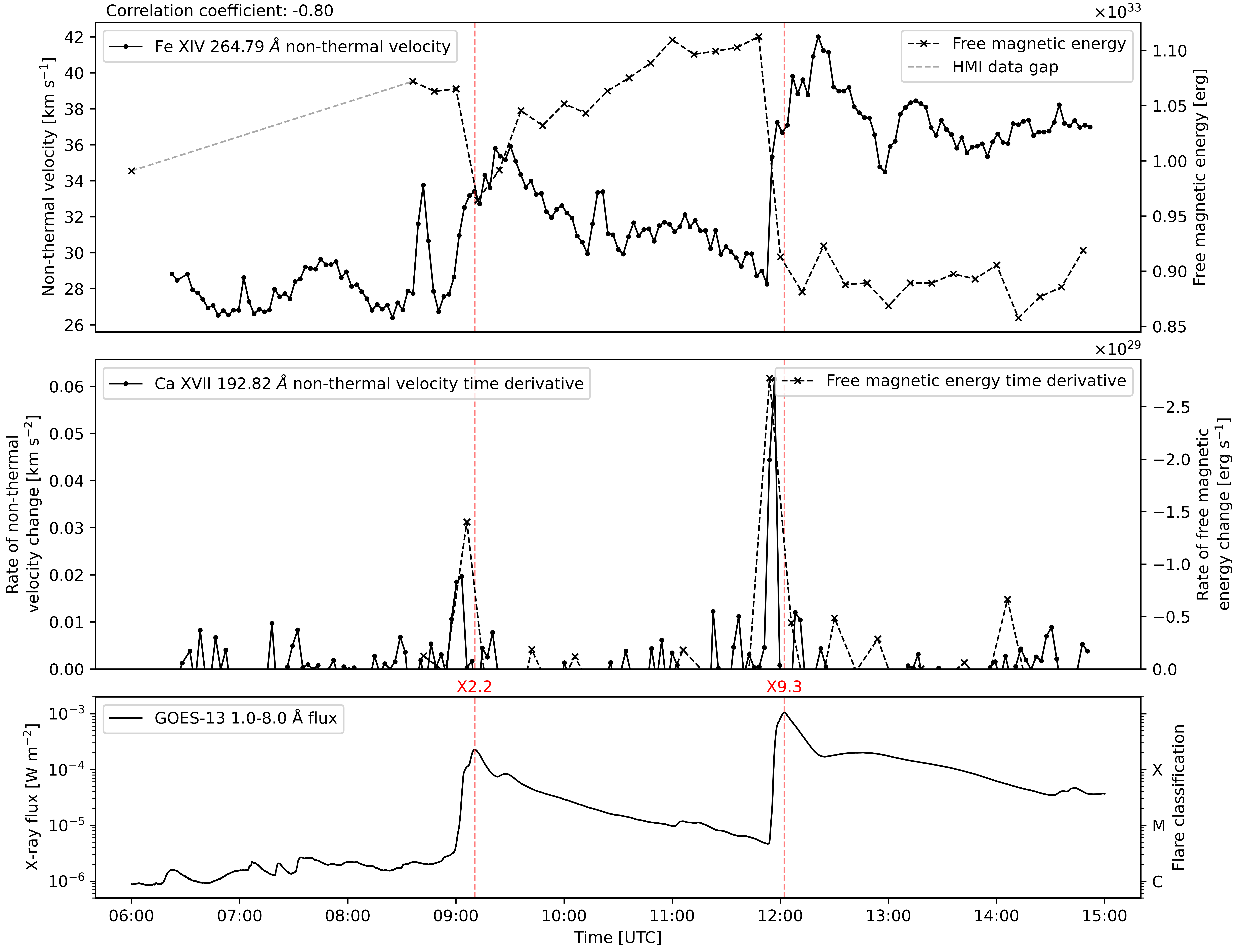}
    \caption{Top panel: The non-thermal velocity time series from 6 September 2017 as derived from the \ion{Fe}{14}~\SI{264.79}{\angstrom} emission line widths as observed by Hinode/EIS (solid line; left axis). Also plotted is the free magnetic energy as estimated using potential and non-potential field modelling based on observations performed by SDO/HMI for SHARP 7115 (dashed line; right axis). An HMI data gap at the beginning of the window is highlighted by a grey line. Middle panel: The non-thermal velocity time derivative for the \ion{Ca}{17}~\SI{192.82}{\angstrom} emission line as observed by EIS (solid line; left axis) is plotted alongside the free magnetic energy time derivative, plotted flipped so that negative values are shown at the top (dashed line; right axis). As discussed in the text, the calculated \ion{Ca}{17}~\SI{192.82}{\angstrom} is a lower estimate. Bottom panel: The soft X-ray flux as observed by the GOES-13 spacecraft between \SIrange[range-phrase = { and }]{1.0}{8.0}{\angstrom} is plotted. A red vertical line is added across all the panels at the time of peak soft X-ray flux to show the flaring times and annotated to identify the flare classification.}
    \label{fig:temporal}
\end{figure*}

Figure~\ref{fig:temporal} shows the evolution of non-thermal velocity and free magnetic energy~-~at 3-minute and 12-minute cadence respectively~-~in active region NOAA~12673 over a 9-hour period encompassing two X-class flares. The times and classifications of the two X-class flares that occurred during this time period are also identified using the peak soft X-ray flux data from GOES-13, these being confined X2.2 and eruptive X9.3 flares at 09:10:25~UTC and 12:02:13~UTC respectively. The magnitudes of magnetic energy calculated by our study are in agreement with those reported by \cite{Fleishman2020DecayFlare}. The Pearson product-moment correlation coefficient was calculated for each time series pair \citep{Bravais1846AnalysePoint}, and found to be $-0.80$ between the \ion{Fe}{14}~\SI{264.79}{\angstrom} non-thermal velocity and free magnetic energy, as seen in the top panel of Figure~\ref{fig:temporal}.


As seen in the top panel of Figure~\ref{fig:temporal}, the level of free magnetic energy in the extrapolated field drops during flaring. This drop is unresolved temporally and takes place within one data point. At the same time, the observed non-thermal velocity rises during flaring, this time being temporally resolved for the first flare and partially temporally resolved for the second. This respective behaviour is strongly negatively correlated. The \ion{Fe}{14}~\SI{264.79}{\angstrom} is shown as it provided a representative time series for the entire 9-hour window, not just during flaring. The \ion{Fe}{14}~\SI{274.20}{\angstrom} emission line results in a similar time series with a slightly stronger negative correlation.

The middle panel of Figure~\ref{fig:temporal} reveals several key features. Firstly, it establishes the temporal coincidence of the increase in non-thermal velocity, the decrease in free magnetic energy, and the peak in soft X-ray flux. The \ion{Ca}{17}~\SI{192.82}{\angstrom} emission line is shown because it is particularly responsive to large flares. Although more noisy, a similar proportional trend is observed in the \ion{Fe}{14}~\SI{264.79}{\angstrom} and \ion{Fe}{14}~\SI{274.20}{\angstrom} emission lines. Secondly, these time series exhibit a proportional response to flaring. The ratio of the peak rate of free magnetic energy decrease to non-thermal velocity increase (\ion{Ca}{17}~\SI{192.82}{\angstrom}) is around \SI{3.5e29}{\erg\second\per\centi\meter} for both flares. This proportional behaviour is not unique to \ion{Ca}{17} and is also evident in the other \ion{Fe}{14} lines we studied. These time series are generated using the entire field of view of \gls{eis} and the entire \gls{nlfff} simulation volume. When considering only the free magnetic energy within the \gls{eis} field of view, a similar correlation is present.

\begin{figure}
    \centering
    \includegraphics[width=.5\textwidth]{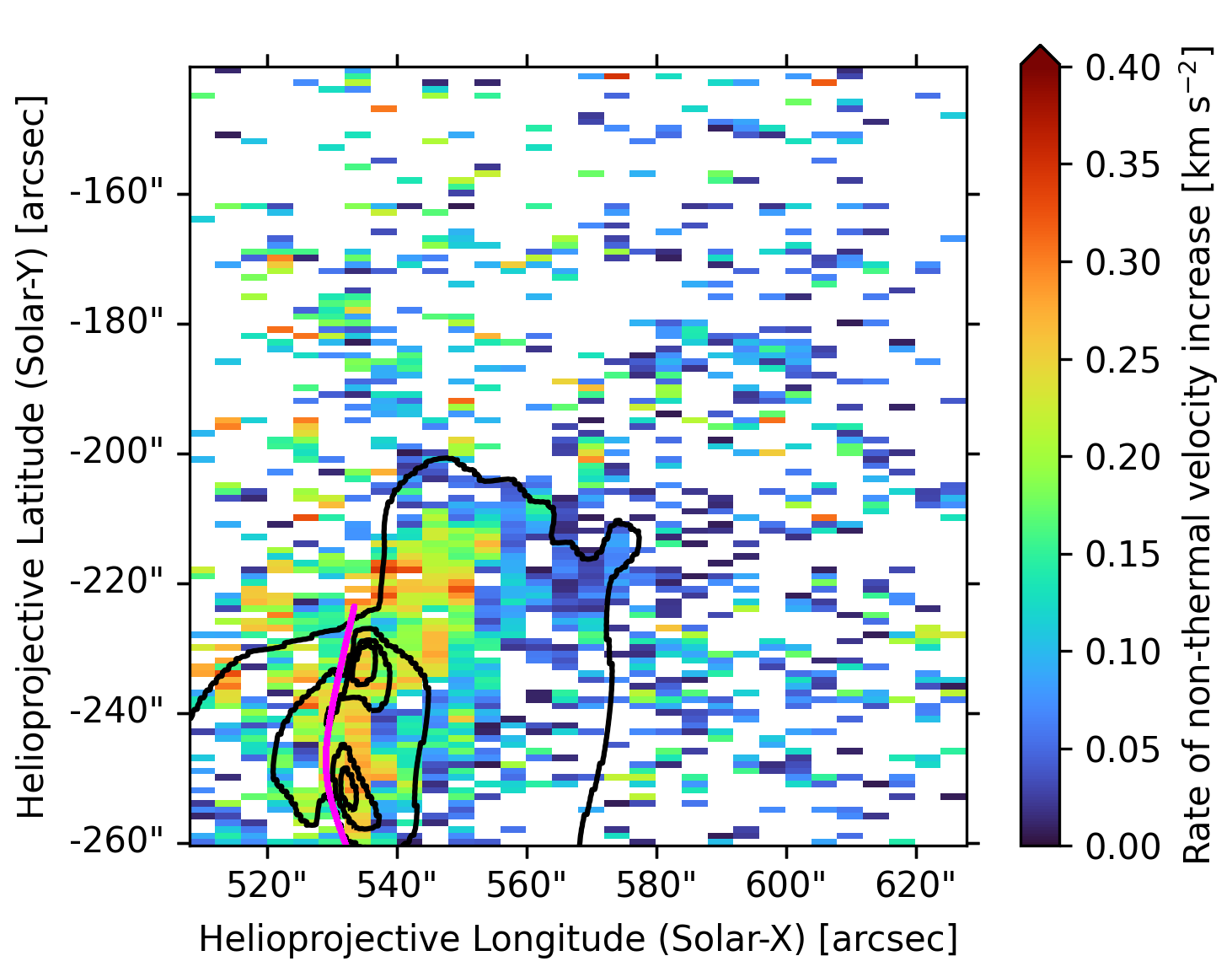}
    \caption{Spatially-resolved rate of non-thermal velocity increase in active region NOAA~12673 on 6 September 2017 between 11:52 and 11:55~UTC as calculated using the \ion{Fe}{14}~\SI{264.79}{\angstrom} emission line. Overplotted is the rate of free magnetic energy decrease between 11:48 and 12:00~UTC, with contours between \mbox{\SIrange[range-phrase = { and }]{-5e5}{-1e5}{\erg\per\second\per\centi\meter\squared}}. The polarity inversion line is illustrated in pink. White pixels represent those either showing a non-thermal velocity decrease, or where the fitting of spectral data was not possible.}
    \label{fig:spatial}
\end{figure}

After identifying the time period of interest for the X9.3 flare, we spatially resolved the respective increases and decreases in non-thermal velocity and free magnetic energy. Figure~\ref{fig:spatial} shows the spatially-resolved time derivative of non-thermal velocity between 11:52 and 11:55~UTC and the spatially-resolved time derivative of free magnetic energy between 11:48 and 12:00~UTC. This respective increase and decrease are positively correlated.

\begin{figure}
    \centering
    \includegraphics[width=.5\textwidth]{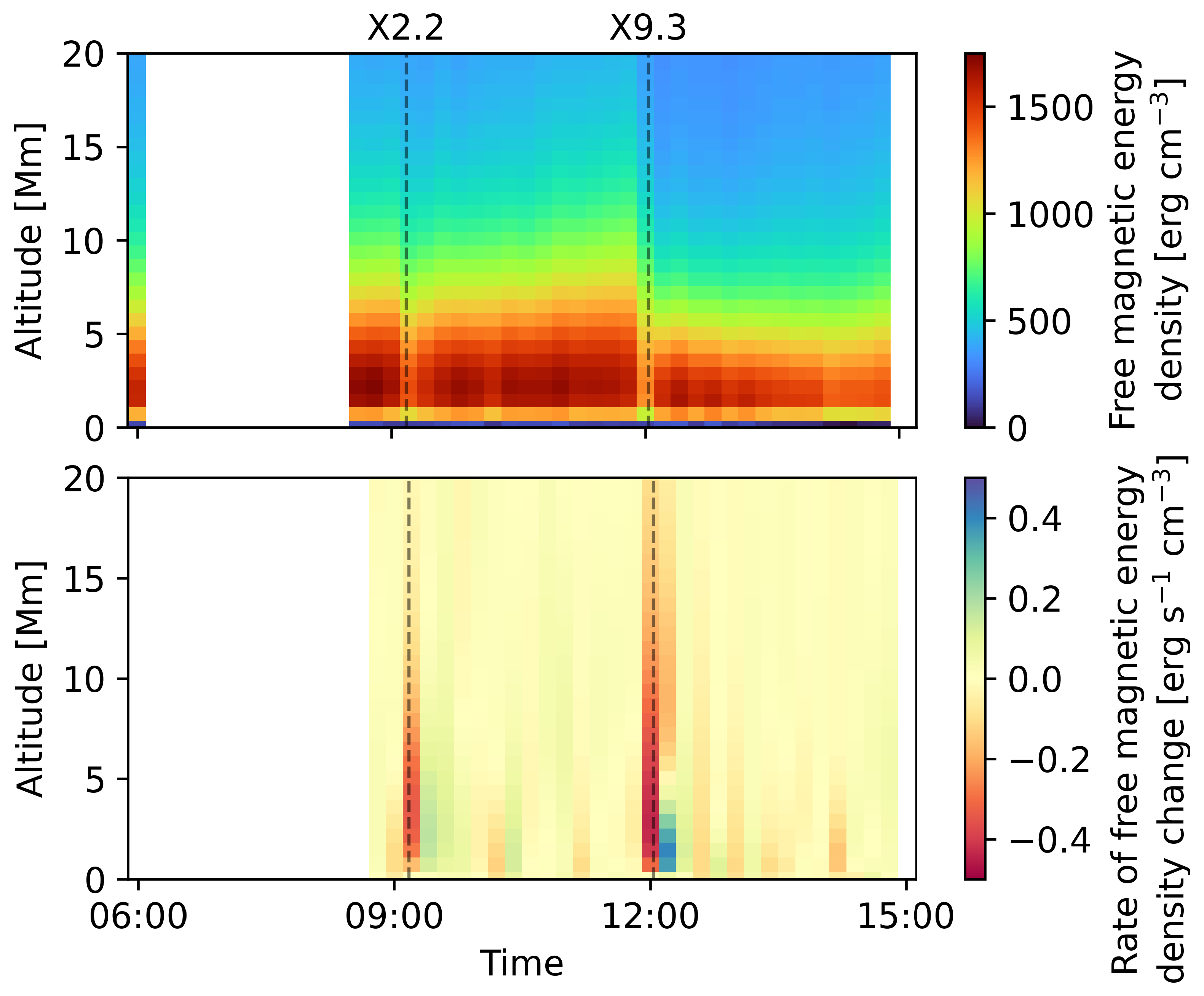}
    \caption{Free magnetic energy density distribution in height (top) and its time derivative (bottom), with the flare times and classifications identified, for 6 September 2017}.
    \label{fig:height}
\end{figure}

Additionally, we calculated the free magnetic energy density distribution in height, and its time derivative, to identify the altitudes at which changes in the free magnetic energy occurred. This is shown in Figure~\ref{fig:height}. The building of the pre-eruptive structure is known to happen prior to an eruptive event and rises before the main eruption \citep{Zhang2001OnFlares,Sterling2005Slow-RiseOnset}. While the \gls{hmi} data gap means no information can be presented prior to the confined X2.2 flare, prior to the eruptive X9.3 flare, the free magnetic energy is seen to increase in magnitude and altitude. This coincides with observations of the corona made at \SI{193}{\angstrom} using \gls{aia}, which also show a slow-rising structure during this period. The time derivative plot shows the peak rate of free magnetic energy decrease to be at around \SI{5}{\mega\metre} in altitude.

\begin{figure}
  \gridline{\fig{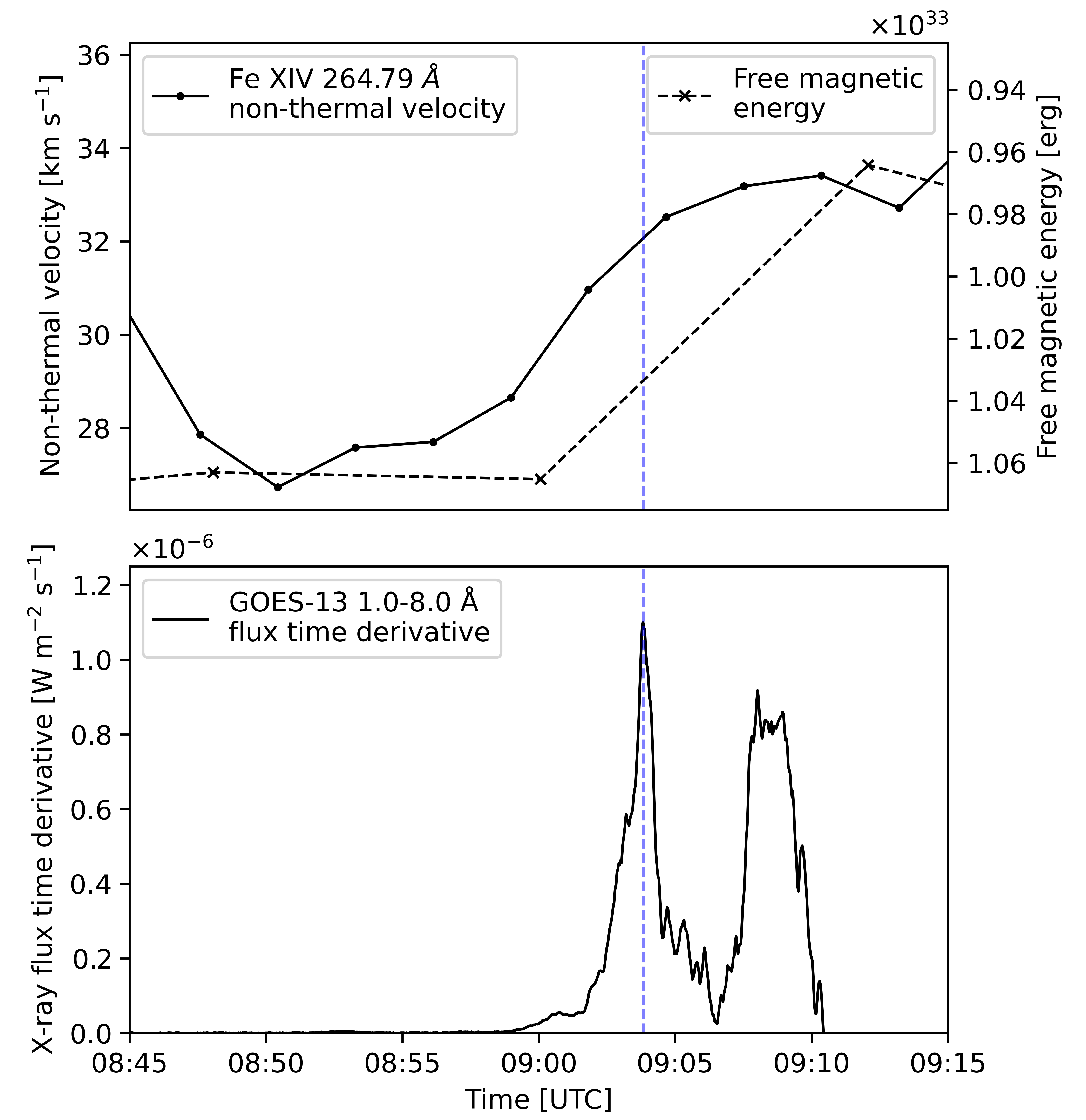}{0.5\textwidth}{(a) Non-thermal velocity, free magnetic energy, and soft X-ray flux time derivative for the X2.2 flare.}}
  \gridline{\fig{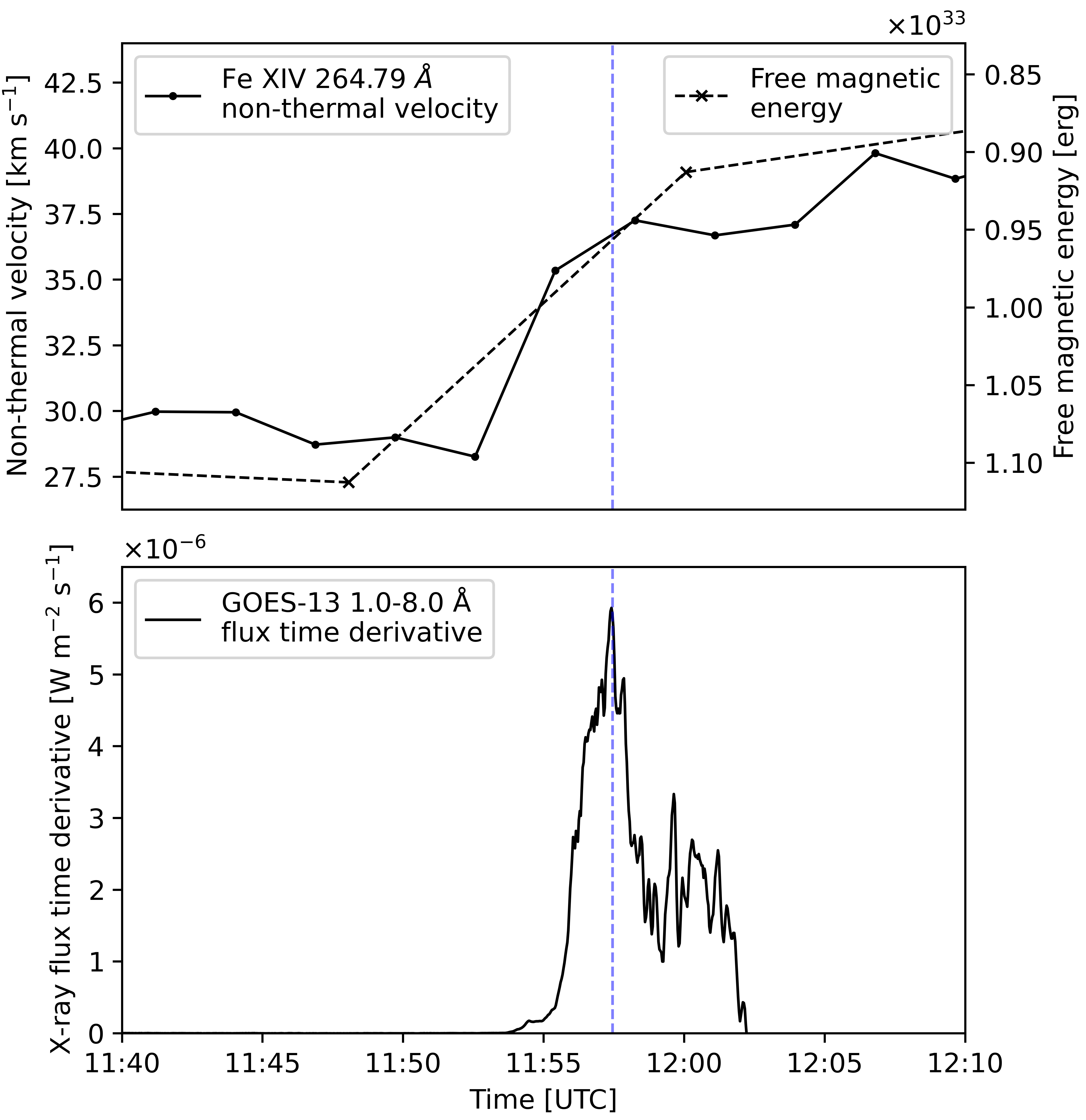}{0.5\textwidth}{(b) Non-thermal velocity, free magnetic energy, and soft X-ray flux time derivative for the X9.3 flare.}}
  \caption{Top panels: The \ion{Fe}{14}~\SI{264.79}{\angstrom} non-thermal velocity (solid line; left axis) and free magnetic energy plotted flipped with lower values at the top (dashed line; right axis). Bottom panels: The time derivative of the GOES soft X-ray flux. A blue vertical line indicates the peak time of the soft X-ray flux time derivative.}
  \label{fig:hxr}
\end{figure}

In order to better understand the chromospheric response during the observed flares, and given the absence of direct hard X-ray data covering both flares, we employ the Neupert effect, which posits a correlation between the time-integrated hard X-ray emission and the soft X-ray emission \citep{Dennis1993TheFlares,Veronig2002InvestigationModel}, as a proxy to investigate the potential role of chromospheric evaporation in driving non-thermal velocities. Figure \ref{fig:hxr} presents the temporal evolution of the non-thermal velocity, alongside the time derivative of the 1--8 {\AA} soft X-ray flux as observed by the GOES-13 spacecraft.

\section{Discussion}

The primary result of this study is the strong temporal and spatial coupling between increases in non-thermal velocity and decreases in free magnetic energy during the two large flares considered. These changes occur within minutes of the rise and peak in soft X-ray flux, indicative of the solar flare energy release process: a rapid conversion of magnetic energy into kinetic and thermal energy of the plasma, in line with the standard flare model. This behaviour is also broadly in agreement with the energy budget of flares described by \cite{Aschwanden2017GlobalEjections}.

As discussed in \cite{Polito2019CanFlares} and references found herein, the possible causes of excess line broadening during solar flares include: the superposition of unresolved flows with various Doppler-shifted components; Alfvén wave propagation accelerating ions perpendicular to the magnetic field; departures from ionisation equilibrium as the result of high temperatures; isotropic turbulence.

\subsection{Superposition of unresolved flows}

Using modelling of superposed Doppler-shifted flows, \cite{Polito2019CanFlares} concluded that this mechanism fails to explain the broadening observed by IRIS \citep[Interface Region Imaging Spectrometer;][]{DePontieu2014TheIRIS} during an X-class flare. Such flows would be produced as the result of chromospheric evaporation, a secondary response to the primary energy release, and typically on timescales of approximately 100~seconds \citep{Ning2012ChromosphericFlares}. The close timing of free magnetic energy decrease and non-thermal velocity increase found in our study is in broad agreement with these findings, although our limited observational cadence, 12 minutes for the former and 3 minutes for the latter, makes it difficult to provide conclusive support. The Neupert effect suggests that the non-thermal electron bombardment, typically observed in hard X-rays, heats the chromosphere, leading to chromospheric evaporation filling the coronal loops with hot plasma which results in the enhanced soft X-ray emission. Therefore, the time derivative of the soft X-ray flux can serve as an indicator of the non-thermal electron precipitation in the absence of hard X-ray data.
In both the X2.2 and X9.3 flares, the peak in the soft X-ray flux derivative, indicative of the maximum rate of chromospheric evaporation, is seen in Figure~\ref{fig:hxr} to occur after observed increases in non-thermal velocity. We, therefore, conclude that the increase in the observed non-thermal velocity is more closely connected to the initial energy release in the corona than to the response to energy deposition in the chromosphere.

\subsection{Alfvén wave propagation}

The spatial, temporal and spectral resolution of the \gls{eis} observations in our study make it challenging to draw any conclusions about the presence of Alfvén waves as the main cause of line broadening. However, we note that \cite{DePontieu2022ProbingHeating} demonstrate that the inclusion of Alfvén waves in flare simulations can result in increased line broadening of magnitudes similar to those measured in this work and on similar timescales.

\subsection{Departures from ionisation equilibrium}

Using \gls{eis} observations of highly ionised Fe lines in flares, \cite{Kawate2016DeparturePlasmas} found some evidence of departures from ionisation equilibrium in a small number of pixels (approximately 1\%), from which they concluded that equilibrium holds in most cases for \gls{eis} exposures. While our observations include lines formed at lower temperatures than those studied by \cite{Kawate2016DeparturePlasmas}, and while departures cannot be completely ruled out, we consider non-thermal broadening from this effect unlikely.

\subsection{Isotropic turbulence}

The question of the origin and presence of turbulence is closely linked to the conditions that are favourable to the onset and evolution of magnetic reconnection and/or instability, and the release of free magnetic energy. However, which comes first remains a major open question. \cite{French2021ProbingDynamics} found evidence supporting the development of the tearing mode instability prior to the increase in excess line broadening followed by a rapid increase in line broadening and evolution of the energy spectrum to a turbulence dominated regime, something supported by simulations \citep{Dong2018RoleTurbulence,Tenerani2020SpectralReconnection}. For the events studied here, we observe a gradual increase in line broadening followed by a rapid increase coincident with the drop in free magnetic energy for the first X-class flare, and a coincident increase in line broadening and a decrease in free magnetic energy for the second. With the caveat that our temporal resolution is low compared to typical impulsive phase timescales, we suggest that the close inverse relationship found in our work is consistent with the scenario of a turbulent cascade in response to free magnetic energy release.

\cite{Harra2013TheRopes} observed differences between eruptive and confined events in the pre-flare enhancement of non-thermal velocity, with a confined event showing enhancement only in the flaring region and eruptive events additionally showing enhancement at footpoints and close to or above the loop regions. The height of the behaviour we observed in the plasma is difficult to ascertain relative to the height of the behaviour we observed in the magnetic field given only line of sight observations near to vertically above the flare were made by \gls{eis}. However, we find our observation of a concentrated non-thermal velocity enhancement in the eruptive flare, in the same region as a free magnetic energy drop, to be broadly consistent with the findings of \cite{Harra2013TheRopes}.

\subsection{Conclusions}

In conclusion, our study reveals a strong temporal and spatial correlation between increases in non-thermal velocity and decreases in free magnetic energy during solar flares, consistent with the standard flare model of reconnection-driven energy conversion. Our findings contribute to the understanding of the complex interplay between magnetic fields and plasma dynamics in solar flares, and are consistent with the suggestion that Alfvén wave propagation and isotropic turbulence are more likely to be responsible for non-thermal line broadening.

\section*{Acknowledgements}

We are thankful to the referee for the comments and suggestions that helped to improve the manuscript. The authors would like to thank Julia Thalmann for her support of the original SOLARNET proposal, and also thank her and Manu Gupta for their invaluable insights into the \acrshort{nlfff} technique. This research was supported by STFC PhD Studentship number ST/X508858/1 (J.M.), the European Union’s Horizon 2020 research and innovation programme under grant agreement No. 824135 (SOLARNET) (J.M., R.J., A.M.V.), Hinode Ops Continuation 2022-25 grant number ST/X002063/1 (S.M., D.B.), Solar System Consolidated Grant 2022-25 ST/W001004/1. (S.M., H.R.) and Solar Orbiter EUI Operations grant number ST/X002012/1 (D.B., H.R.). \gls{nlfff} extrapolations and data analysis were performed on the Vienna Scientific Cluster (VSC)\footnote{\href{https://vsc.ac.at/}{https://vsc.ac.at/}}.

\bibliography{main}{}
\bibliographystyle{aasjournal}



\end{document}